\documentclass[prl,twocolumn,superscriptaddress,floatfix,nopacs]{revtex4}
\usepackage{graphicx,amsfonts,amssymb,amsmath,hyperref,enumerate}

\newif\ifhyper
\hypertrue
\ifhyper
\hypersetup{
   citecolor = {red},
   colorlinks = {true}, 
   linkcolor = {blue},
   urlcolor = {blue} 
}
\fi

\newcommand{\beq}{\begin{equation}}
\newcommand{\eeq}{\end{equation}}
\newcommand{\beqa}{\begin{eqnarray}}
\newcommand{\eeqa}{\end{eqnarray}}
\newcommand{\ket} [1] {\vert #1 \rangle}
\newcommand{\bra} [1] {\langle #1 \vert}
\newcommand{\braket}[2]{\langle #1 | #2 \rangle}

\def\bra#1{\langle#1\vert}
\def\ket#1{\vert#1\rangle}

\def\Longarrow{\protect\@lra}
\def\@lra{\relbar\joinrel\relbar\joinrel\relbar\joinrel%
          \relbar\joinrel\rightarrow}

\begin{document}

\title{Fine-Grained Tensor Network Methods}

\author{Philipp Schmoll}
\affiliation{Institute of Physics, Johannes Gutenberg University, 55099 Mainz, Germany}

\author{Saeed S. Jahromi}
\affiliation{Donostia International Physics Center, Paseo Manuel de Lardizabal 4, E-20018 San Sebasti\'an, Spain}

\author{Max H\"ormann}
\affiliation{Chair for Theoretical Physics I, FAU Erlangen-N\"urnberg, Germany}

\author{Matthias M\"uhlhauser}
\affiliation{Chair for Theoretical Physics I, FAU Erlangen-N\"urnberg, Germany}

\author{Kai Phillip Schmidt}
\affiliation{Chair for Theoretical Physics I, FAU Erlangen-N\"urnberg, Germany}

\author{Rom\'an Or\'us}
\affiliation{Donostia International Physics Center, Paseo Manuel de Lardizabal 4, E-20018 San Sebasti\'an, Spain}
\affiliation{Ikerbasque Foundation for Science, Maria Diaz de Haro 3, E-48013 Bilbao, Spain}
\affiliation{Multiverse Computing, Pio Baroja 37, 20008 San Sebasti\'an, Spain}

\begin{abstract}
We develop a strategy for tensor network algorithms that allows to deal very efficiently with lattices of high connectivity. The basic idea is to \emph{fine-grain} the physical degrees of freedom, i.e., decompose them into more fundamental units which, after a suitable coarse-graining, provide the original ones. Thanks to this procedure, the original lattice with high connectivity is transformed by an isometry into a simpler structure, which is easier to simulate via usual tensor network methods. In particular this enables the use of standard schemes to contract infinite 2d tensor networks -- such as Corner Transfer Matrix Renormalization schemes --  which are more involved on complex lattice structures. We prove the validity of our approach by numerically computing the ground-state properties of the ferromagnetic spin-1 transverse-field Ising model on the 2d triangular and 3d stacked triangular lattice, as well as of the hard-core and soft-core Bose-Hubbard models on the triangular lattice. Our results are benchmarked against those obtained with other techniques, such as perturbative continuous unitary transformations and graph projected entangled pair states, showing excellent agreement and also improved performance in several regimes.

\end{abstract}

\maketitle

\emph{Introduction.-} During the past decade there has been a rapid development of tensor network (TN) states and numerical methods \cite{Orus2014,Orus2019,Ran2017,Biamonte2017,Verstraete2008} for simulating strongly correlated quantum many-body systems. These are mathematical objects which use the knowledge about the amount and structure of entanglement in quantum many-body states in order to reproduce the state accordingly. TN methods use such objects as \emph{ans\" atze} to simulate quantum lattice systems in different regimes, and have been remarkably successful  \cite{Bauer2012,Corboz2010a,Corboz2010,Corboz2014a,Corboz2014,Jahromi2018a,Jahromi2018,Jahromi2019,Sadrzadeh2016}. Inspiringly, TN states also show up in other disciplines, such as quantum gravity \cite{Swingle2012}, artificial intelligence \cite{Stoudenmire2018,Huggins2019} and even linguistics \cite{Gallego2017}.

Despite being extremely versatile, TNs are not free from limitations, though. The most obvious one is the ability to capture the expected \emph{structure} of entanglement in the TN ansatz, i.e., to incorporate the correct scaling of the entanglement entropy.
The \emph{amount} of entanglement is also a limitation itself, where one of the key parameters of the TN, the so-called \emph{bond dimension}, may be just too large to simulate the system at hand when there is too much entanglement in the quantum state. In addition to these limitations, one also has to deal with \emph{geometric} bottlenecks. For instance, the simulation of a triangular lattice with projected entangled pair states (PEPS) \cite{Verstraete2006,Verstraete2008,Orus2014,Orus2014a} would naively imply tensors with six bond indices, if we were to use one tensor per lattice site. As such, handling tensors with so many indices quickly becomes computationally expensive for numerical simulations. The same problem also arises for higher-dimensional systems, where high-connectivity lattices are quite usual. This is a serious issue, since such large-connectivity lattices are usually linked to exotic phases of  matter such as quantum spin liquids \cite{Savary2017,Balents2010,Yan2011,Jahromi2016,Jahromi2017}. 

Here we propose a physically motivated strategy to solve this problem, which on top is remarkably efficient and accurate. The idea is to break down the physical degrees of freedom into ``smaller" pieces, i.e., to \emph{fine-grain the lattice}. This can be done at the expense of introducing a set of \emph{fine-graining isometries}. The key advantage is that \emph{the fine-grained lattice is easily amenable to TN methods}. Unlike other proposals of TN methods for high-connectivity lattices \cite{Corboz2010a,Corboz2012a,Corboz2014,Niesen2018a,Jahromi2018a,Jahromi2018,Jahromi2019}, our approach preserves the correct geometric structure of the system, thus being better-suited in terms of the entanglement structure. In what follows we explain the approach and use it to compute ground-state properties of the ferromagnetic spin-1 transverse-field Ising model on the triangular and 3d stacked triangular lattice, as well as of the hard-core and soft-core Bose-Hubbard models on the triangular lattice. We benchmark the results against those obtained with perturbative continuous unitary transformations (pCUTs) \cite{Knetter2000,Knetter2003,Coester2015} and graph projected entangled pair states (gPEPS) \cite{Jahromi2019}, showing excellent agreement and also improved performance in several regimes.

\emph{Method.-} Our approach is based on a simple yet powerful idea: split the physical degrees of freedom into smaller, more fundamental entities which, when coarse-grained, reproduce the original physical ones. In other words, fine-grain the local Hilbert spaces at each site. 

Before proceeding any further let us give a practical example. Imagine that we have a spin-$1$ particle. As is well known, this can always be understood as two spin-$1/2$ particles which are projected into their spin-1 subspace in the coupled basis. Mathematically, since for $SU(2)$ irreps one has $1/2 \otimes 1/2 = 0 \oplus 1$, what we do is to project out the singlet part with spin $0$ and keep the triplet with spin $1$. In this way, we constructed a spin-$1$ out of two spins-$1/2$. But we can also consider the procedure the other way around: we fine-grain a spin-$1$ into two spins-$1/2$ by using the appropriate ``inverse" projector, i.e.~a \emph{fine-graining isometry}, which in this particular case is the Clebsch-Gordan coefficient $\braket{1/2, 1/2, m_1, m_2}{1/2, 1/2, 1, m}$ with $m_{1,2} = \pm 1/2$ and $m = -1, 0, +1$, using the standard notation $\braket{j_1, j_2, m_1, m_2}{j_1, j_2, 1, m}$. 

\begin{figure}
	\centering
	\includegraphics[width=1\linewidth]{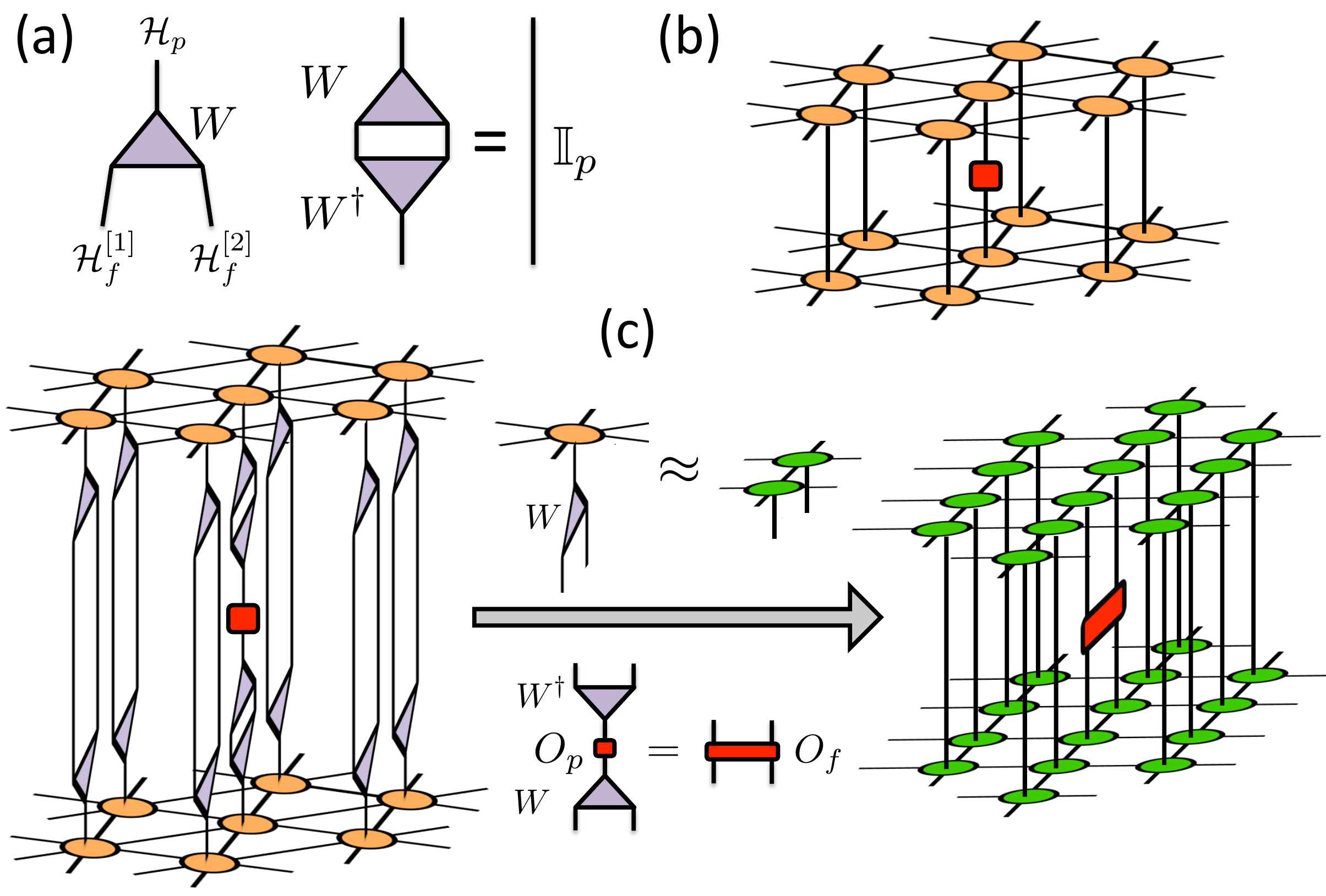}
	\caption{(Color online) (a) Isometry $W$ projects the fine-grained Hilbert spaces $\mathcal{H}^{[1]}_f$ and $\mathcal{H}^{[2]}_f$ into the physical space $\mathcal{H}_p$. The isometry verifies $W^\dagger W = {\mathbb I_p}$, with ${\mathbb I_p}$ the identity in the physical space; (b) Expectation value of an one-site operator for a 2d PEPS on a triangular lattice; (c) By introducing resolutions of the identity $W^\dagger W$ at every site, we can rewrite the expectation value in terms of a fine-grained two-site operator and fine-grained PEPS tensors on a 2d square lattice.}
	\label{fig1}
\end{figure}

The idea above is generalized as follows: a physical degree of freedom described by a Hilbert space $\mathcal{H}_p$ can be understood as the coarse-grained space of some other fine-grained Hilbert spaces $\mathcal{H}_{f}^{[1]}$ and $\mathcal{H}_{f}^{[2]}$ via some isometry $W$, i.e., 
\beq
W : \mathcal{H}_{f}^{[1]} \otimes \mathcal{H}_{f}^{[2]} \longrightarrow \mathcal{H}_p, 
\eeq
with $W = \sum_{i f_1 f_2} W^i_{f_1 f_2} \ket{f_1} \ket{f_2} \bra{i}$. In TN language, the 3-index tensor $W^i_{f_1 f_2}$ coarse-grains the indices $f_1$ and $f_2$ into $i$. Seen in reverse, the physical index $i$ is fine-grained into indices $f_1$ and $f_2$ by the isometric tensor $W^i_{f_1 f_2}$. Since $W$ is an isometry, it implies that $W^\dagger W = {\mathbb I_p}$, with ${\mathbb I_p}$ the identity in the physical Hilbert space $\mathcal{H}_p$, see Fig.~\ref{fig1}.a. Let us remark that we considered here the case of two fine-grained Hilbert spaces, but the idea can be easily generalized to having more than two. In fact, the whole isometry $W$ could even have a TN structure itself (as in, e.g., the multiscale entanglement renormalization ansatz (MERA) \cite{Vidal2007b}), if required. Generically, the isometry can also be understood in the language of entanglement branching operators \cite{Harada2018}.

Next, we apply this fine-graining to the physical degrees of freedom of many-body systems with high-connectivity, which allows us to simplify the underlying lattice structure and therefore make them more amenable to TN simulation methods. Let us consider, without loss of generality, the case of a triangular lattice. As shown in Fig.~\ref{fig1}.b-c, fine-graining every site maps the triangular lattice into a \emph{square} lattice. The key point is to realize that, in such a scenario, \emph{operators on the triangular lattice are mapped to operators on the fine-grained square lattice via the isometry W}, as shown in Fig.\ref{fig1}.c. For instance, for an one-site operator $O_p$ acting on one site of the physical lattice, one has 
\beq
O_f = W O_p W^\dagger
\eeq
with $O_f$ the corresponding operator on the fine-grained lattice. In the case of the triangular lattice that we are discussing, this maps a one-site operator $O_p$ on the triangular lattice to a two-site operator $O_f$ on the square lattice. In general, for a fine-graining isometry involving $n$ fine-grained Hilbert spaces, an $m$-body operator on the original lattice is mapped to an $(n \times m)$-body operator in the fine-grained one. 

Our method can thus be summarized in three steps: 
\begin{enumerate}
\item{Find an isometry $W$ that reduces the connectivity of the lattice after fine-graining.}
\item{Use $W$ to map all operators involved in the TN algorithm to their fine-grained versions.}
\item{Run the TN algorithm on the fine-grained lattice using the fine-grained operators.}
\end{enumerate}

The mapping between lattices preserves locality inasmuch the isometry $W$ is local. This implies, for instance, that local expectation values in the original lattice may also be mostly local in the fine-grained one. Notice also that, at the level of TN optimization and calculation of local expectation values, one can \emph{fully} operate in the fine-grained space only, see Fig.~\ref{fig1}.c for an example. 

\begin{figure}
	\centering
	\includegraphics[width=\linewidth]{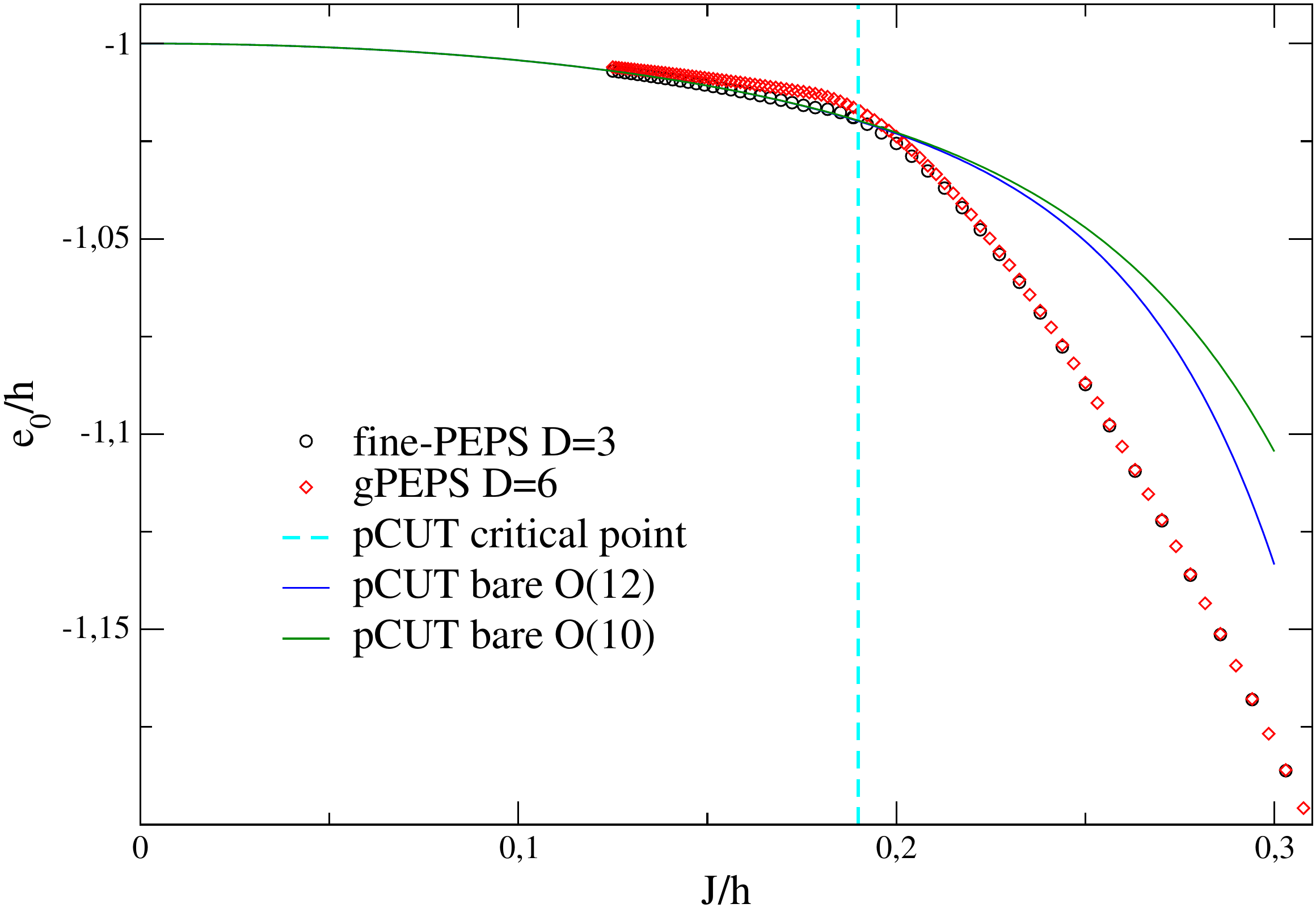}
	\caption{(Color online) Ground-state energy per site for the ferromagnetic spin-$1$ quantum Ising model on the triangular lattice, as computed by fine-PEPS (circles), gPEPS (diamonds), and pCUTs (solid lines). The vertical dashed line refers to the critical point \mbox{$(h/J)_{\rm c}^{\rm pCUT}=0.1898(1)$} from extrapolating the one-particle pCUT gap.}
	\label{fig3}
\end{figure}

A number of practical considerations are in order. First, the isometry $W$ is a new degree of freedom that enters the TN algorithm. It could be optimized following a MERA-like procedure, yet another option is to fix it to some reasonable choice and optimize over the tensors of the fine-grained TN. This choice is not unique and moreover it is also reasonable to think that some isometries may work better than others in practice depending on the symmetries of the system. Generally, an isometry that splits the physical Hilbert space symmetrically seems to be beneficial (e.g., a decomposition of $1 = 0 \otimes 1$ is valid but unbalanced). Second, interaction terms in the fine-grained Hamiltonian may become of slightly longer range. For instance, for a Hamiltonian with nearest-neighbour interactions on the triangular lattice, one gets interactions that span over four sites in the fine-grained square lattice. Third, and as we said above, more complicated isometries are also possible, even with an internal TN structure. Further discussion about the details of the method and the relevant tensor updates can be found in Ref.~\cite{supp,Vidal2003}.

\emph{Numerical results.-} In order to benchmark the validity of our approach we computed the ground-state properties of several models on the triangular lattice for a unit cell of $2 \times 2$ tensors. For this, we used fine-graining together with the infinite-PEPS algorithm \cite{Phien2015,Orus2009} on the square lattice with a $2 \times 4$ unit cell and simple update, also for four-body interactions, and computed expectation values with Corner Transfer Matrix (CTM) techniques \cite{Orus2009,Corboz2014a,Corboz2010a}.

The first model that we considered is the spin-$1$ ferromagnetic quantum Ising model in a transverse field, described by the Hamiltonian \cite{Powalski2013} 
\beq
H = -J \sum_{\langle i, j \rangle} \sigma^{[i]}_x \sigma^{[j]}_x - h \sum_i \sigma^{[i]}_z,
\label{eq:TFIM_Hamiltonian}
\eeq
with $\sigma^{[i]}_\alpha$ the $3\times3$ spin-one matrix at site $i$, $J > 0$ the ferromagnetic interaction strength, and $h$ the magnetic field. It realizes a polarized phase for small $J/h$ and a symmetry-broken ordered phase for large $J/h$ separated by a second-order phase transition in the 3d Ising universality class. The location of the critical point can be estimated precisely by the pCUT series of the one-particle gap in the polarized phase using Dlog Pad\'{e} extrapolation \cite{Guttmann1989} which yields $(J/h)_{\rm c}^{\rm pCUT}=0.1898(1)$ or equivalently in the inverse unit \mbox{$(h/J)_{\rm c}^{\rm pCUT}=5.269(3)$} \cite{supp,Kos2016}.

\begin{figure}
	\centering
	\includegraphics[width=\linewidth]{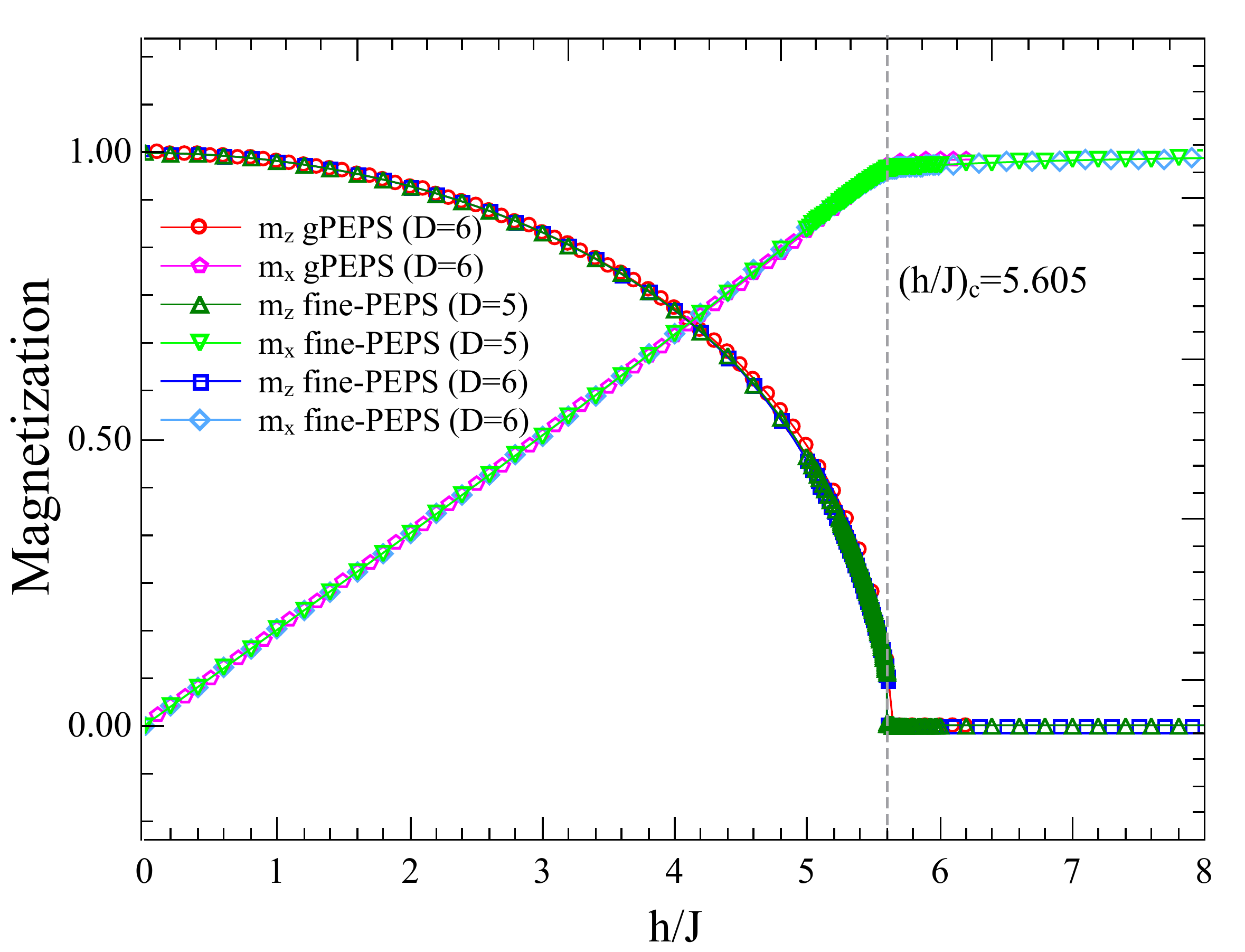}
	\caption{(Color online) Longitudinal and transverse magnetization per site for the ferromagnetic spin-$1$ quantum Ising model on the triangular lattice, as computed by fine-PEPS and gPEPS.}
	\label{fig4}
\end{figure}

For the fine-PEPS we choose to fine-grain each spin-$1$ into two spins-$1/2$ via an isometry that equals a Clebsch-Gordan coefficient, \mbox{$W_{m_1, m_2}^m = \braket{1/2, 1/2, m_1, m_2}{1/2, 1/2, 1, m}$.} In Fig.~\ref{fig3} we show the ground-state energy per site computed by fine-graining (fine-PEPS) with PEPS bond dimension $D=3$, as well as using gPEPS with $D=6$ and pCUT up to $O(12)$ in the high-field expansion in $J/h$. Remarkably, even for a small bond dimension $D=3$, the agreement of fine-PEPS with pCUT for $J/h\leq (J/h)_{\rm c}^{\rm pCUT}$ within the polarized phase and with gPEPS for large $J/h$ inside the symmetry-broken ordered phase is almost perfect. In Fig.~\ref{fig4} we also plot longitudinal and transverse magnetizations as computed by fine-PEPS and gPEPS, also in excellent agreement, and with an approximate quantum critical point of $(h/J)_{\rm c}^{\rm fine-PEPS} \approx 5.605$. Notice that the critical point obtained by the two tensor network methods deviates from the pCUT result $(J/h)_{\rm c}^{\rm pCUT}$. This is, however, due to the simple update, which does not make use of the full environment when updating the tensors. Simulations with the full environment would improve the accuracy close to criticality, shall this be required.

\begin{figure}
	\centering
	\includegraphics[width=\linewidth]{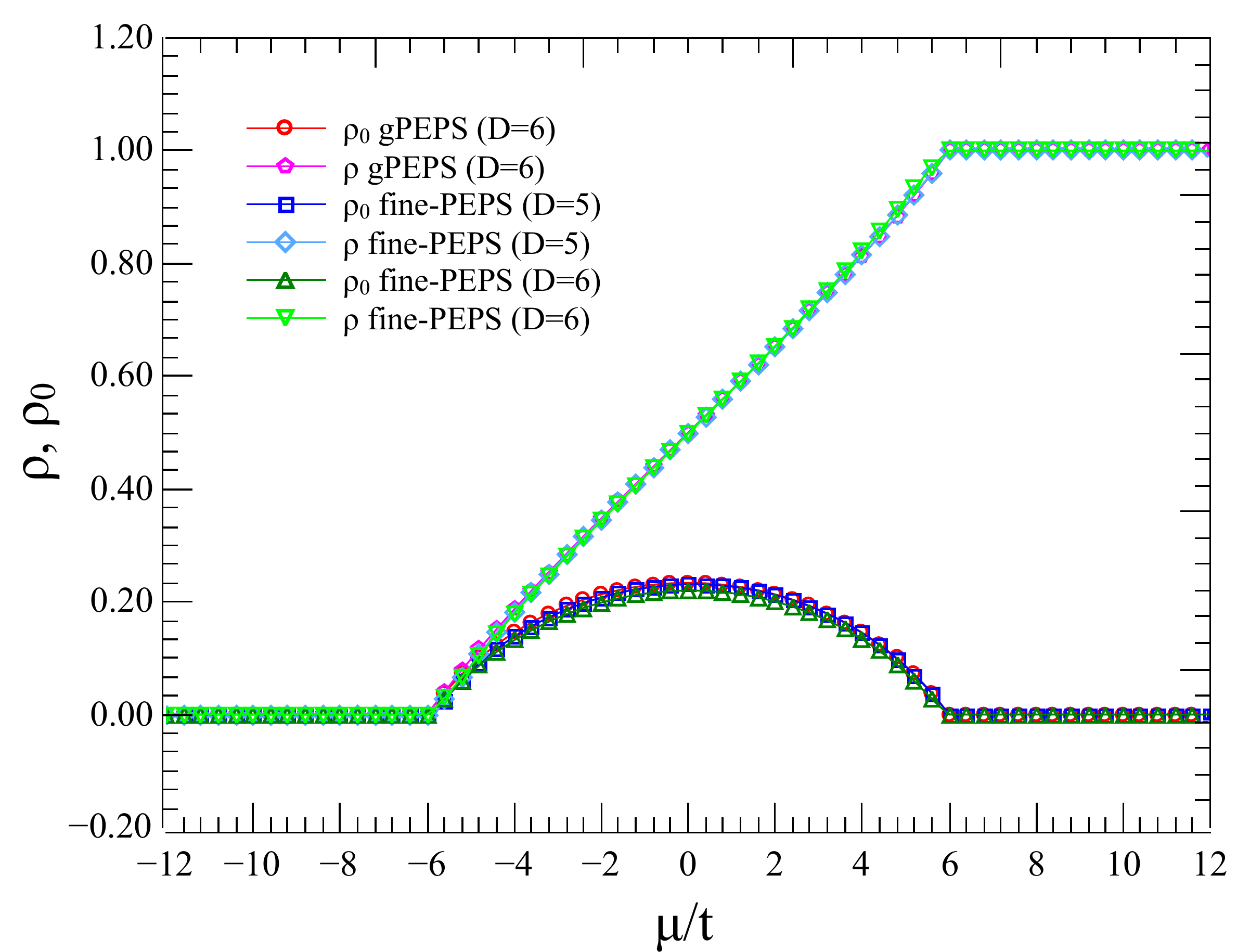}
	\caption{(Color online) Particle density and condensate fraction for the hard-core Bose-Hubbard model on the triangular lattice, for $t=1$, as computed by fine-PEPS and gPEPS.}
	\label{fig5}
\end{figure}	

Furthermore, we simulated the Bose-Hubbard model on the triangular lattice, described by the Hamiltonian \cite{Kshetrimayum2019,Wang2013} 
\beq
H = -t \sum_{\langle i, j \rangle} \left( a^\dagger_i a_j + h.c. \right) + \frac{U}{2} \sum_i n_i \left( n_i - 1\right) - \mu \sum_i n_i, 
\eeq
with $a_j, a_j^\dagger$ and $n_j = a_j^\dagger a_j$ respectively being bosonic annihilation, creation and number operators at site $j$, $t$ the hopping strength, $U$ the on-site density-density interaction, and $\mu$ the chemical potential. 

In the hard-core limit $U\rightarrow \infty$, where individual sites are either empty or occupied by one boson, this model realizes two exact gapped Mott phases with density zero and one as well as an intermediate gapless superfluid phase. The phase transitions at $(\mu/J)_{\rm c}=\pm 6$ between the Mott and superfluid phases can be determined exactly by first-order perturbation theory for the one-particle gap of the two Mott phases \cite{supp}. Technically, we fine-grain every hard-core boson into two hard-core bosons via an isometry with non-zero coefficients \mbox{$W_{0,0}^0 = 1, W_{1,0}^1 = W_{0,1}^1 = 1/\sqrt{2}$}. Thus, if the physical site is occupied, then the hard-core boson can be on either of the fine-grained sites. In Fig.~\ref{fig5} we show our numerical results for the particle density $\rho = \langle a^\dagger_j a_j \rangle$ and the condensate fraction $\rho_0 = |\langle a_j \rangle |^2$ for fine-PEPS and gPEPS both up to $D=6$ and with $t=1$, showing excellent agreement in the superfluid and Mott-insulator phases. 

\begin{figure}
	\centering
	\includegraphics[width=\linewidth]{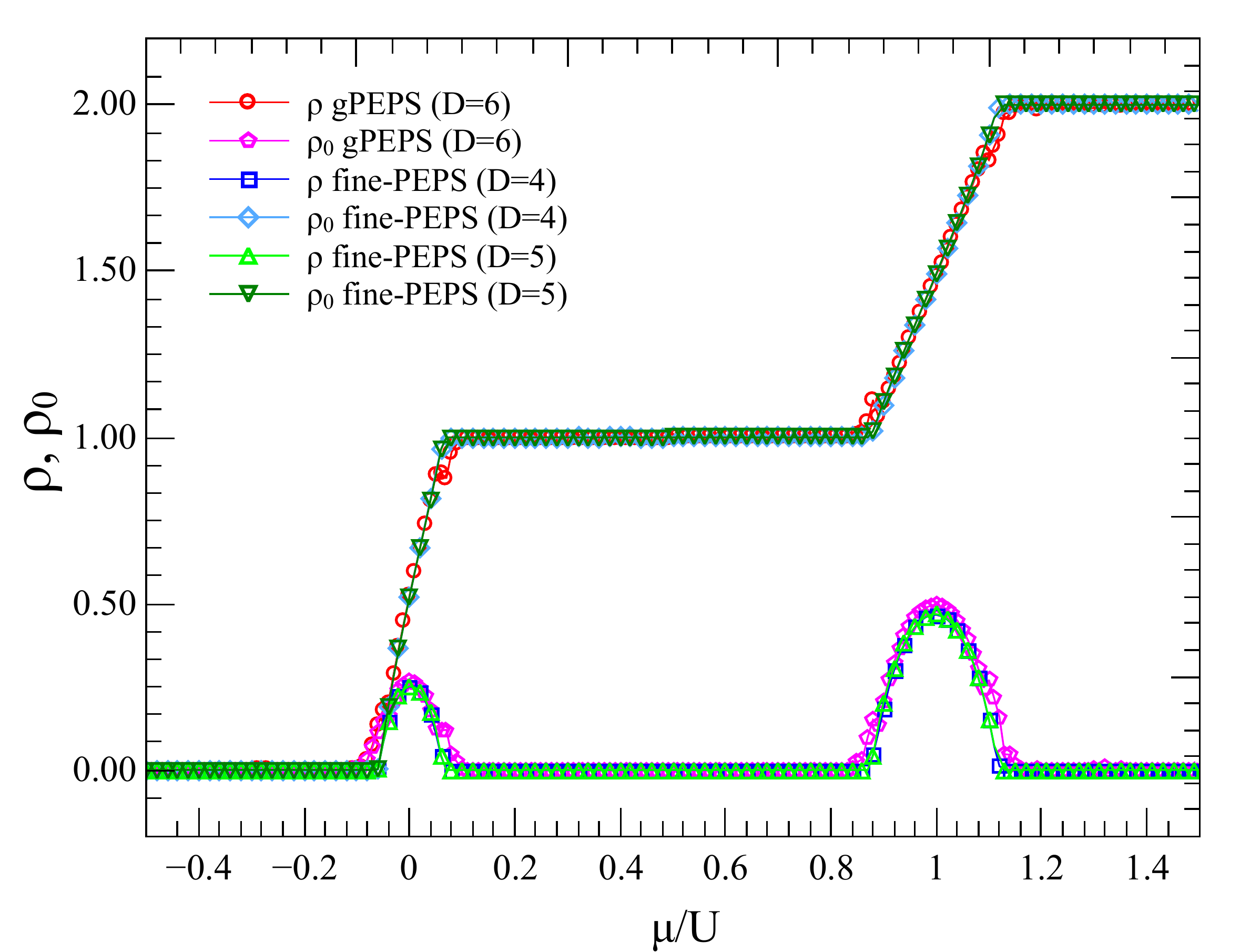}
	\caption{(Color online) Particle density and condensate fraction for the soft-core Bose-Hubbard model on the triangular lattice with up to $2$ bosons per site, for $t=0.01$ and $U=1$, as computed by fine-PEPS and gPEPS.}
	\label{fig6}
\end{figure}

Furthermore, we considered the soft-core case up to two bosons per lattice site so that the ground-state phase diagram consists of three Mott lobes with densities $n\in\{0,1,2\}$ and superfluid phases. The empty ($n=0$) and completely filled ($n=2$) Mott states are again exact eigenstates of the system and the corresponding one-particle gap $\Delta_{n=0}^{\rm p}=-\mu-6t$ (one-hole gap \mbox{$\Delta_{n=2}^{\rm h}=-U+\mu-12t$}) can be calculated exactly \cite{supp}. This is different for the Mott phase with $n=1$ where the hopping term introduces quantum fluctuations. For the fine-PEPS we break down again each local site in terms of two hard-core bosons which, when both occupied, result in a double occupied physical site. For this we use an isometry with non-zero coefficients \mbox{$W_{0,0}^0 = 1, W_{1,0}^1 = W_{0,1}^1 = 1/\sqrt{2}, W_{1,1}^2 = 1$}. The particle density and condensate fraction for soft-core bosons is shown in Fig.~\ref{fig6}, computed by fine-PEPS up to $D=5$ and gPEPS up to $D=6$ with $t=0.01$ and $U=1$, again showing an excellent agreement in all superfluid and Mott-insulating phases. 

\begin{figure}
	\centering
	\includegraphics[width=\linewidth]{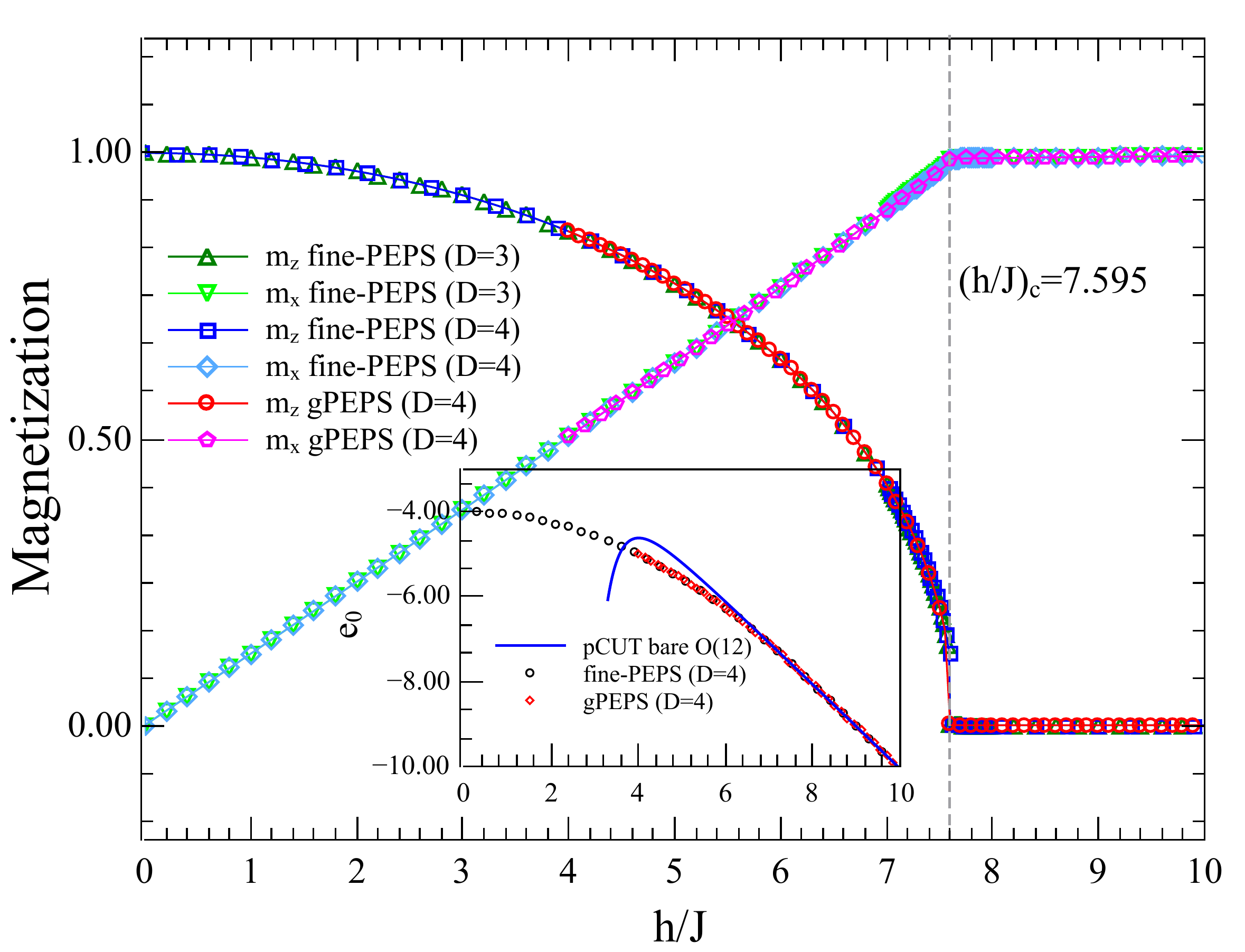}
	\caption{(Color online) Longitudinal and transverse magnetization per site for the ferromagnetic spin-$1$ quantum Ising model on the 3d stacked-triangular lattice, as computed by fine-PEPS and gPEPS. The inset shows the ground-state energy per site of the ITF model obtained with fine-PEPS, gPEPS, and pCUT (bare order 12 is shown).}
	\label{fig7}
\end{figure}

In order to show the potential of our method in higher dimensions we consider Eq.~\ref{eq:TFIM_Hamiltonian} on a 3d stacked triangular lattice (see \cite{supp} for a depiction of the lattice structure). The location of the critical point with expected mean-field exponents can be estimated again precisely by extrapolating the pCUT series of the one-particle gap in the polarized phase which yields \mbox{$(h/J)_{\rm c}^{\rm pCUT}=7.45(1)$} \cite{supp}. Here the local iPEPS tensors have eight virtual indices besides the physical one. Using the same idea of fine-graining the local Hilbert space of the spin-ones, the model is mapped onto a cubic lattice. Importantly, this mapping enables the use of 3d CTM schemes to perform the contraction of the infinite 3d lattice. Due to the reduced importance of quantum fluctuations in 3d we choose however to use the mean-field environment for every local iPEPS tensor in the present simulations. Fig.~\ref{fig7} shows the magnetization as well as the ground-state energy as a function of the magnetic field. We find $(h/J)_{\rm c}^{\rm fine-PEPS} \approx 7.59$ which is in good agreement with that of gPEPS and pCUT.

\emph{Conclusions.-} In this paper we have proposed an efficient approach to deal with lattices of high connectivity in TN methods, by using a fine-graining of the physical degrees of freedom. Under suitable conditions, this fine-graining simplifies the lattice and essentially keeps locality of interactions. After a fine-graining of operators, the approach allows us to apply usual TN methods on simpler lattices in a remarkably efficient way. Most importantly, the fine-graining allows us to use the CTM method for approximating the contraction of the infinite TN, in turn, capturing all quantum correlations into the environment of local tensors which are also essential for the full update iPEPS simulations. This is a huge advancement over other TN methods such as gPEPS which use mean-field environments for calculations of the expectation value of local operators and correlators. We have explained in detail the example of the 2d triangular lattice, which in our approach can be simulated using standard 2d square-lattice PEPS algorithms. Our method has been benchmarked with numerical simulations of the ground state of paradigmatic magnetic and bosonic models in 2d and 3d, with excellent accuracy when compared to other methods such as pCUT and gPEPS. We believe that the approach in this paper will allow to overcome the computational cost associated to simulating lattices of high connectivity, such as the ones typically found for higher dimensional systems and frustrated quantum antiferromagnets and will become an instrumental tool in the discovery of new exotic phases of quantum matter. 
\acknowledgements 
We acknowledge discussions with A. Haller, A. Kshetrimayum and M. Rizzi. We also acknowledge DFG funding through GZ OR 381/3-1 as well as GZ SCHM 2511/10-1. 

\bibliography{references}{}
\bibliographystyle{apsrev4-1}

 \end{document}


\title{Fine-Grained Tensor Network Methods\\ 
\textit{Supplemental Material}}

\author{Philipp Schmoll}
\affiliation{Institute of Physics, Johannes Gutenberg University, 55099 Mainz, Germany}

\author{Saeed S. Jahromi}
\affiliation{Donostia International Physics Center, Paseo Manuel de Lardizabal 4, E-20018 San Sebasti\'an, Spain}

\author{Max H\"ormann}
\affiliation{Chair of Theoretical Physics I, FAU Erlangen-N\"urnberg, Germany}

\author{Matthias M\"uhlhauser}
\affiliation{Chair of Theoretical Physics I, FAU Erlangen-N\"urnberg, Germany}

\author{Kai Phillip Schmidt}
\affiliation{Chair of Theoretical Physics I, FAU Erlangen-N\"urnberg, Germany}

\author{Rom\'an Or\'us}
\affiliation{Donostia International Physics Center, Paseo Manuel de Lardizabal 4, E-20018 San Sebasti\'an, Spain}
\affiliation{Ikerbasque Foundation for Science, Maria Diaz de Haro 3, E-48013 Bilbao, Spain}
\affiliation{Multiverse Computing, Pio Baroja 37, 20008 San Sebasti\'an, Spain}

\maketitle

This supplementary material contains details about the simple update for the triangular fine-grained iPEPS, the calculation of expectation values of local operators within the fine-grained iPEPS, the 3d stacked triangular lattice, the method of perturbative continuous unitary transformations, and exact considerations for the Mott phases of the Bose-Hubbard model.

\section{Simple Update for the triangular fine-grained iPEPS}
\setcounter{figure}{0}
\setcounter{equation}{0}

In the simple update of the fine-grained triangular lattice we use the standard infinite time-evolving block decimation (iTEBD) \cite{Vidal2003} procedure to determine the tensors that represent the ground state of the model. One step of the simple update includes all links in the triangular lattice, and the procedure is repeated with decreasing Trotter steps until convergence of the singular values is reached. Choosing an $L_x$ times $L_y$ unit cell on the triangular lattice we have to update $3 L_x L_y$ links, which is done in the resulting $L_x$ times $2L_y$ unit cell on the square lattice. Due to the splitting of the physical sites the update of every link now involves four iPEPS tensors instead of only two. In order to lower the computational cost we decompose the input tensors so that all virtual indices that are not affected by the update are separated \cite{Phien2015,Jahromi2019}, and the simple update can be performed more efficiently on the reduced tensors. After the Suzuki-Trotter gate has been applied the resulting tensor is decomposed using an singular-value decomposition (SVD), which yields the updated singular values on the particular link as well as the tensors $U$ used to restore tensors A and B on the left, and $V^\dagger$ used to restore tensors C and D on the right. Notice that the singular values $S_{AB}$ and $S_{CD}$ between tensors A, B and tensors C, D respectively are updated too, even though their links are introduced artificially due to the splitting.
\begin{figure}[ht]
  \centering
  \includegraphics[width = 0.7\textwidth]{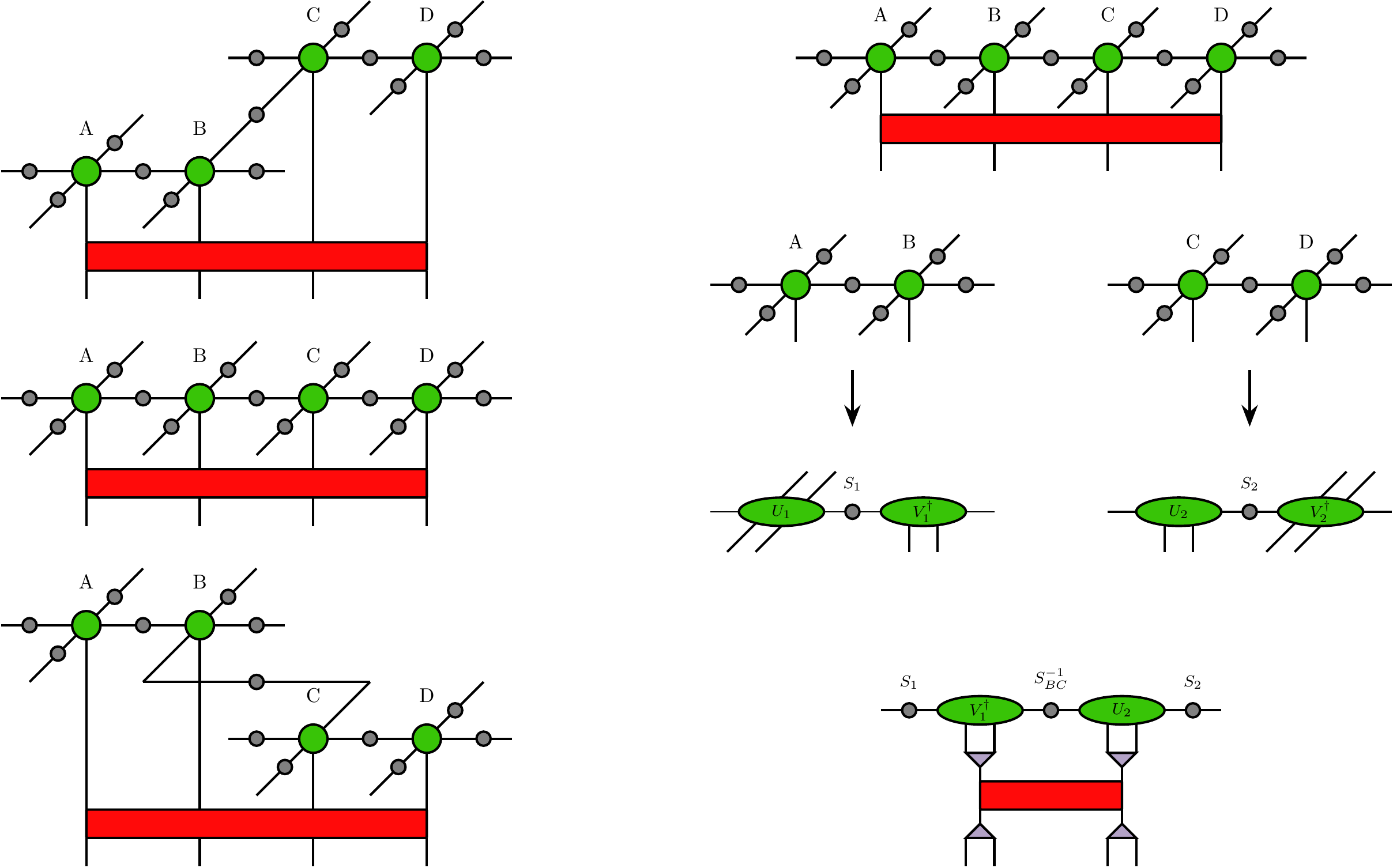}
  \caption{Simple update in the fine-PEPS algorithm. The three different type of links in the triangular lattice and their update using fine-grained tensors (left) and the decomposition of involved iPEPS tensors and the application of the TEBD gate on the reduced tensors (right). The example shows the update of a horizontal link in the network corresponding to the middle figure on the right.}
  \label{fig:SM_SimpleUpdate_1}
\end{figure}
Eventually this procedure is a simple update for the triangular lattice with extra steps to maintain the fine-grained structure, however we can exploit the fine-graining for the computation of expectation values in the next section.

\section{Expectation values of local operators}

\begin{figure}[ht]
  \centering
  \includegraphics[width = 0.7\textwidth]{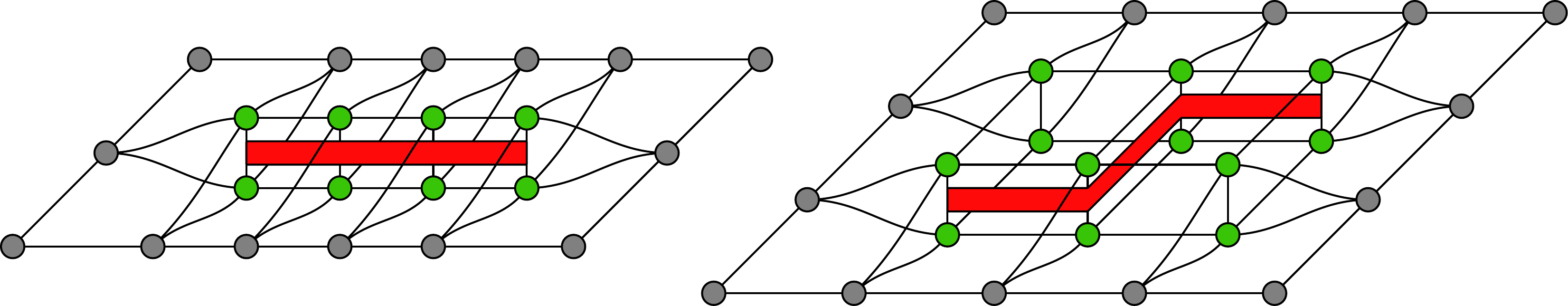}
  \caption{Computation of two-site observables like the Hamiltonian in the triangular lattice translates to evaluating four-site operators in the fine-grained square lattice. Here we show the expectation values for horizontal and one type of diagonal links, see also Fig.~\ref{fig:SM_SimpleUpdate_1}.}
  \label{fig:SM_ExpectationValue_1}
\end{figure}

Computing the ground state wave function using simple update as described above is not too expensive even for high-connectivity tensors. However computing expectation values becomes more difficult, also because of the geometric structure of the triangular lattice. It is convenient to resort to the square lattice for which one can define an effective environment easily. The environment tensors are computed in a directional CTM algorithm \cite{Orus2009,Corboz2014a,Corboz2010a} for an arbitrary $L_x \times L_y$ unit cell, the iPEPS tensors are absorbed into all lattice directions iteratively until the environment is converged. The CTM tensors then represent the contraction of the infinite two-dimensional square lattice.\\
For the computation of expectation values we can use the CTM tensors as an effective environment for the iPEPS tensors in the unit cell. Since we can directly compute expectation values in the fine-grained lattice the operator support will be doubled, e.g. a one-site operator on the triangular lattice will become a two-site operator in the square lattice, and so on. Computing the energy per link in the triangular lattice is then done by computing the expectation value of a four-site operator in the square lattice, which is obtained by fine-graining the physical indices of the regular two-site Hamiltonian. For a more efficient contraction of the resulting tensor networks in Fig.~\ref{fig:SM_ExpectationValue_1} we can decompose the four-body gate into two parts that act on both fine-grained sites.
The norm of the quantum states is computed similarly, just without the operators. For one-site operators like the particle density $\rho = \langle a_j^\dagger a_j \rangle$ we only use two two fine-grained sites and their respective environment tensors.

\section{The 3d stacked triangular lattice and other applications}
\begin{figure}[ht]
  \centering
  \begin{minipage}{0.3\textwidth}
    \includegraphics[width = 0.8\textwidth]{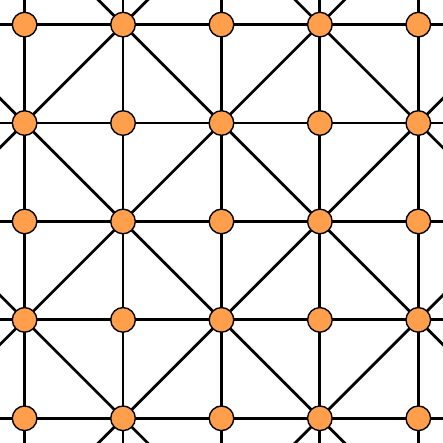}
  \end{minipage}
  \hfill
  \begin{minipage}{0.3\textwidth}
    \includegraphics[width = 0.8\textwidth]{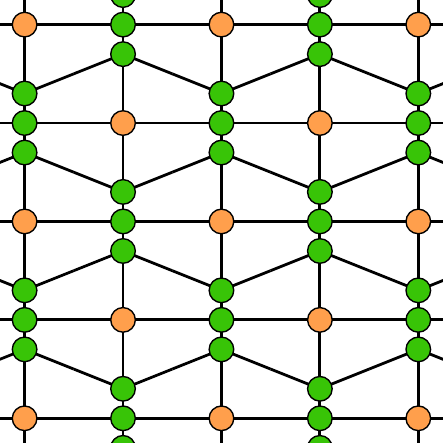}
  \end{minipage}
  \hfill
  \begin{minipage}{0.38\textwidth}
    \includegraphics[width = 0.9\textwidth]{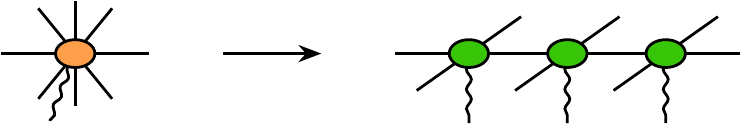}
  \end{minipage}
  \caption{A different form of a triangular lattice for which the fine-graining procedure can be applied. Here the lattice sites that connect to eight neighbours are fine-grained using a 1-to-3 isometry which results in a square lattice.}
  \label{fig:fineGraining_Example_2}
\end{figure}

In the manuscript we considered the 3d stacked triangular lattice as an example of the application of fine-graining technique to simulation of a 3d lattice model. The unit cell of the lattice is easily incorporated in gPEPS as a graph that connects the vertices, i.e. the local iPEPS tensors. The fine-PEPS algorithm uses a $(N_x,N_y,N_z) = (2,2,2)$ tensor unit cell on the 3d stacked triangular lattice, corresponding to a $(N_x,N_y,N_z) = (2,4,2)$ tensor unit cell on the fine-grained 3d counterpart which is equivalent to a cubic-like lattice.
\begin{figure}[ht]
  \centering
  \begin{minipage}{0.49\textwidth}
    \includegraphics[width = 0.65\textwidth]{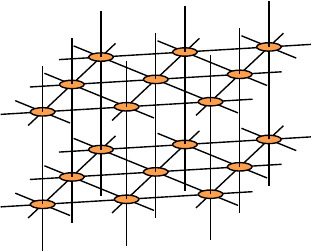}
  \end{minipage}
  \hfill
  \begin{minipage}{0.49\textwidth}
    \includegraphics[width = 0.9\textwidth]{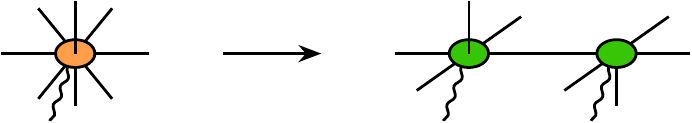}
  \end{minipage}
  \caption{The infinite 3d stacked triangular lattice used for both tensor network and pCUT simulations of the ITF model on the left. On the right we show the decomposition of the nine-index tensors using an isometry as described in the main text, resulting in two six-index tensors on the fine-grained cubic-like lattice. The physical indices are shown as curly lines.}
  \label{fig:fineGraining_3d_1}
\end{figure}
For the fine-PEPS algorithm we apply the fine-graining procedure proposed in the paper, so that the nine-index tensor on the stacked triangular lattice is decomposed into two six-index tensors on the cubic-like lattice as shown in Fig.~\ref{fig:fineGraining_3d_1}. For the computation of observables we used the mean-field environment instead of the full environment computed by a CTM procedure. This is reasonable because the quantum correlations are less important in 3d compared to lower dimensions.\\
Finally let us provide yet another lattice for which our fine-graining method can be applied. For the particular triangular lattice in Fig.~\ref{fig:fineGraining_Example_2} on the left every site that connects to eight neighbours is fine-grained using a 1-to-3 isometry. It is therefore mapped to a regular square lattice for which all the techniques discussed in the paper can be applied.

\section{Comparison between fine-PEPS and gPEPS}
In this section we elaborate on the benefits of the fine-PEPS over other established TN methods like gPEPS. First of all the comparison to gPEPS is reasonable, because gPEPS is one of the most general simple update simulation tools for arbitrary lattices. A comparison of 2d and 3d TN simulations is therefore possible. Our fine-PEPS method is computationally slightly less efficient for the simple update due to the overhead of maintaining the fine-grained structure, the overall complexity is however similar. Using the mean-field environment for the computation of expectation values for fine-PEPS and gPEPS is therefore equivalent.\\
The main advantage of fine-PEPS is however the mapping to a simpler lattice structure, which readily enables the use of standard algorithms to contract the infinite 2d lattice such as the CTM method. This is not only relevant for computing accurate expectation values but also is essential for the full update iPEPS algorithm which is able to capture all quantum correlations by taking the whole environment into account~\cite{Phien2015}. The computational cost for the CTM procedure can be compared between fine-PEPS and a fine-grained version of gPEPS. Here each gPEPS tensor is split just before the CTM procedure, i.e. after running the simple update to obtain a ground state approximation. Due to this late splitting, the bond dimension between two fine-grained gPEPS tensors is significantly enlarged. Using e.g. an SVD to split the gPEPS tensor, the connecting bond dimension is $\mathcal O(dD^3)$ for the 2d triangular, and $\mathcal O(dD^4)$ for the 3d stacked triangular lattice. The resulting fine-grained tensors then have an average (virtual) bond dimension of
\begin{align}
  D_\text{av}^{2d} = \frac{3D + dD^3}{4} \hspace{1.0cm} \text{or} \hspace{1.0cm} D_\text{av}^{3d} = \frac{4D + dD^4}{5}
\end{align}
which makes the CTM procedure much less efficient compared to fine-PEPS. Note that the CTM procedures can also be implemented on lattices of higher connectivity (and not only on the square lattice), however the efficiency here is naturally also greatly reduced.

\section{Perturbative Continuous Unitary Transformation}
The high-order linked-cluster expansions are realized with the help of perturbative continuous unitary transformations (pCUTs) \cite{Knetter2000,Knetter2003}. In the following we describe its generic aspects and refer to the literature for further details.

One can always rewrite any Hamiltonian ${\cal H}$ exactly as
\begin{equation}
\label{Eq:Hami}
{\cal H}={\cal H}_0+\sum_j \lambda_j {\cal V}^{(j)}\quad ,
\end{equation}
where the $\lambda_j$ are the perturbative parameters and the unperturbed part $\mathcal{H}_0$ is diagonal in appropriate supersites. For the high-field expansion in the spin-1 transverse-field Ising model as well as for the density $n=1$ Mott phase in the softcore Bose-Hubbard model we use single sites as supersites. In both cases one can express $\mathcal{H}_0$ in appropriate units as 
\begin{eqnarray}\label{h_0_q}
\mathcal{H}_0 &=& E_0+\mathcal{Q} \quad ,
\end{eqnarray}
where $E_0$ denotes a constant and $\mathcal{Q}$ is a counting operator of local excitations. This decomposition of $\mathcal{H}_0$ is always possible, since the local spectra of the supersites is equidistant in all considered cases.

Supersites interact via the perturbation \mbox{$\mathcal{V}\equiv\sum_j\lambda_j\mathcal{V}^{(j)}$}. In this work always pairs of nearest-neighbor supersites on the triangular lattice are linked by the perturbation. As a consequence of Eq.~\eqref{h_0_q}, one can rewrite Eq.~\eqref{Eq:Hami} as
\begin{equation}
\label{Eq:Hami_final}
{\cal H}={\cal H}_0+ \sum_{n=-N}^N \hat{T}_n \quad ,
\end{equation}
so that $[\mathcal{Q},\hat{T}_n]=n\hat{T}_n$. Physically, the operator \mbox{$\hat{T}_n \equiv\sum_j \lambda_j \hat{T}^{(j)}_n$} corresponds to all processes where the change of energy quanta with respect to $\mathcal{H}_0$ is exactly $n$. The maximal (finite) change in energy quanta is called $\pm N$, which is always $N=2$ in the models considered below.

In pCUTs, Hamiltonian \eqref{Eq:Hami_final} is mapped model-independently up to high orders in perturbation to an effective Hamiltonian $\mathcal{H}_\text{eff}$ with $[\mathcal{H}_{\rm eff},\mathcal{Q}]=0$. The general structure of $\mathcal{H}_{\rm eff}$ is then a weighted sum of operator products $\hat{T}_{n_1}\cdots \hat{T}_{n_k}$ in order $k$ perturbation theory. The block-diagonal $\mathcal{H}_\text{eff}$ conserves the number of quasi-particles (qp). This represents a major simplification of the quantum many-body problem, since one can treat each quasi-particle block, corresponding only to a few-body problem, separately, e.g.~, the 0qp-block is given by a single matrix element representing the ground-state energy in all considered cases. 

The more demanding part in pCUTs is model-dependent and corresponds to a normal-ordering of $\mathcal{H}_\text{eff}$ for which the explicit processes of ${\cal H}_0$ and $\mathcal{V}$ have to be specified. This can be either done via a full graph decomposition in linked graphs using the linked-cluster theorem and an appropriate embedding scheme afterwards \cite{Coester2015} or by calculations on large enough finite clusters, which include all relevant virtual processes. Here we did a full graph decomposition and we concentrated on the 0qp and 1qp pCUT sector, which allows us to extract the ground-state energy per site $e_0$ and the elementary gap $\Delta$.

\subsection{Extrapolation}
We perform standard DlogPad\'{e} extrapolations for the one-particle gap $\Delta$. We refer to the literature for a general review of this topic, as for example given in Ref.~\onlinecite{Guttmann1989}. Here we give specific information which is relevant for the particular extrapolation we performed in the main body of the manuscript. 

Our series are all of the form
\begin{align}
F(\lambda)=\sum_{k\geq 0}^{k_{\mathrm{max}}} c_k \lambda^k=c_0+c_1\lambda+c_2\lambda^2+\dots c_{k_{\mathrm{max}}}\lambda^{k_{\mathrm{max}}},
\end{align}
with $\lambda\in \mathbb{R}$ and $c_k \in \mathbb{R}$. If one has power-law behavior near a critical value $\lambda_{\rm c}$, the true physical function $\tilde{F}(\lambda)$ close to $\lambda_{\rm c}$ is given by
\begin{align}
\tilde{F}(\lambda)\approx \left(1-\frac{\lambda}{\lambda_{\rm c}}\right)^{-\theta} A(\lambda),
\end{align}
where $\theta$ is the associated critical exponent. If $A(\lambda)$ is analytic at $\lambda=\lambda_{\rm c}$, we can write
\begin{align}
\label{eq:Ftilde}
\tilde{F}(\lambda)\approx \left(1-\frac{\lambda}{\lambda_{\rm c}}\right)^{-\theta}A\Big|_{\lambda=\lambda_{\rm c}}\left(1+\mathcal{O}\left(1-\frac{\lambda}{\lambda_{\rm c}}\right)\right).
\end{align}
Near the critical value $\lambda_{\rm c}$, the logarithmic derivative is then given by
\begin{align}
\tilde{D}(\lambda)&:=\frac{\text{d}}{\text{d}\lambda}\ln{\tilde{F}(\lambda)}\label{dx}\\
&\approx \frac{\theta}{\lambda_{\rm c}-\lambda}\left\{ 1+ \mathcal{O}(\lambda-\lambda_{\rm c})\right\}\nonumber.
\end{align}
In the case of power-law behavior, the logarithmic derivative $\tilde{D}(\lambda)$ is therefore expected to exhibit a single pole at $\lambda\equiv\lambda_{\rm c}$.

The latter is the reason why so-called DlogPad\'{e} extrapolation is often used to extract critical points $\lambda_{\rm c}$ from high-order series expansions. DlogPad\'e extrapolants of $F(\lambda)$ are defined by
\begin{align}
\label{eq:dlogP1}
dP[L/M]_F(\lambda)=\exp\left(\int_{0}^\lambda P[L/M]_{D}\,\,\text{d}\lambda'\right)
\end{align}
and represent physically grounded extrapolants in the case of a second-order phase transition. Here $P[L/M]_{D}$ denotes a standard Pad\'e extrapolation of the logarithmic derivative
\begin{align}
\label{eq:dlogP2}
P[L/M]_{D}:=\frac{P_L(\lambda)}{Q_M(\lambda)}=\frac{p_0+p_1\lambda+\dots + p_L \lambda^L}{q_0+q_1\lambda+\dots q_M \lambda^M}\quad,
\end{align}
with $p_i\in \mathbb{R}$, $q_i \in \mathbb{R}$, and $q_0=1$. Additionally, $L$ and $M$ have to be chosen so that $L+M\leq k_{\mathrm{max}}-1$. Physical poles of $P[L/M]_{D}(\lambda)$ then indicate critical values $\lambda_{\rm c}$. 

For our results we study the possible combinations of the order of the numerator and denominator polynomial $L$ and $M$. We sort them into the families $[n,n-3]$, $[n,n+3]$, $[n,n-2]$, $[n,n+2]$, $[n,n-1]$, $[n,n+1]$, and $[n,n]$ and analyze their convergence.

\subsection{Series expansion Results for the Spin-1 Triangular transverse-field Ising model}

The spin-$1$ ferromagnetic quantum Ising model in a transverse field on the triangular lattice is given by
\beq
\mathcal{H} = -J \sum_{\langle i, j \rangle} \sigma^{[i]}_x \sigma^{[j]}_x - h \sum_i \sigma^{[i]}_z\,, 
\eeq
with $\sigma^{[i]}_\alpha$ the $3\times3$ spin-one matrix at site $i$, $J > 0$ the ferromagnetic interaction strength, and $h$ the magnetic field. 

We perform a high-field linked-cluster expansion using the pCUT method. Indeed, in the limiting case $J=0$ the unperturbed system consists of isolated spin-ones and has an equidistant spectrum. In the following we set $h=1$ which fixes the local energy quanta of a single spin one to unity. Indeed, the local energies of a single spin-one are then given by $\epsilon_m=m$ for the states $|1,m\rangle$ with $m\in\{\pm 1,0\}$. The field term can be written as
\beq
\mathcal{H}_0 = E_0 +\mathcal{Q}\,, 
\eeq
where $E_0=-N$ with $N$ the number of sites is the bare ground-state energy where all spin-ones are in the $m=+1$ configuration and $\mathcal{Q}=\sum_i\left( \hat{n}^{m=0}_i+2\hat{n}^{m=-1}_i\right)$ is the counting operator of local energy quanta with $\hat{n}^{m=0}$ ($\hat{n}^{m=-1}_i$) the occupation number operator of $m=0$ ($m=-1$) configurations on site $i$.
  
The Ising interaction links always nearest-neighbor sites on the triangular lattice  and it changes the number of energy quanta (eigenvalues of $\mathcal{Q}$) by $\pm 2$ or $0$. As a consequence, the Ising perturbation can be expressed as 
\begin{equation}
\label{Eq:Hami_Ising}
{\cal V}=J\left( \hat{T}_{-2} +  \hat{T}_{0} +\hat{T}_{+2}\right)\quad .
\end{equation}
The pCUT method allows now to map, order by order in $J$, this Hamiltonian to an effective one which commutes with $\mathcal{Q}$ so that the number of quasi-particles (qp) is a conserved quantity. 

\begin{figure}[ht]
  \centering \includegraphics[width=0.7\textwidth]{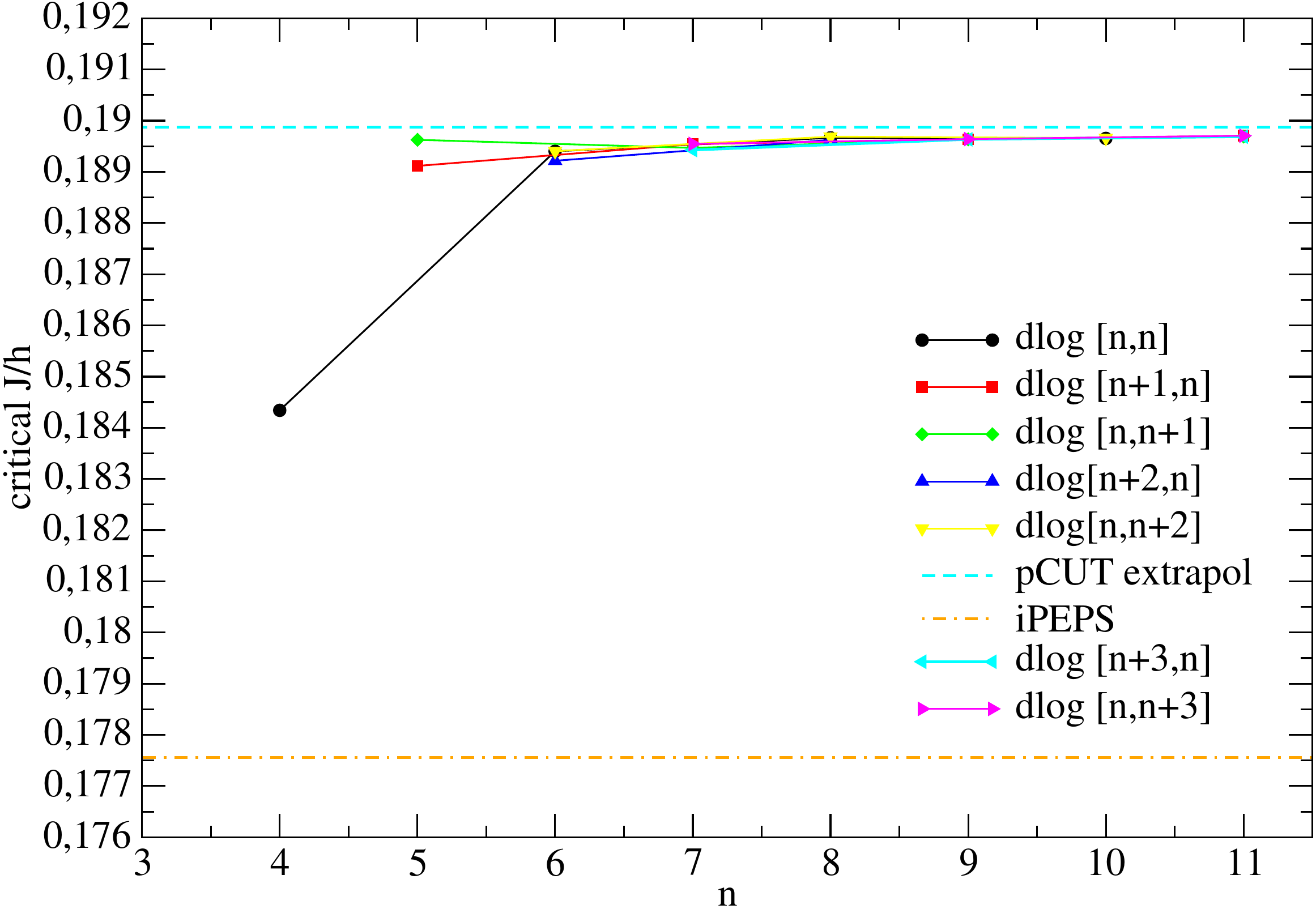}
  \caption{Critical point $(J/h)_{\rm c}^{\rm pCUT}$ of the spin-1 transverse-field Ising model on the triangular lattice as a function of the perturbative order $n$ from Dlog Pad\'{e} extrapolation of the pCUT gap $\Delta$. Connected symbols represent a family of Dlog Pad\'{e} extrapolants $[L,M]$ with $L-M$ fixed. This results in the estimate $(J/h)_{\rm c}^{\rm pCUT}=0.1898(1)$, where the uncertainty reflects the difference of the families in the highest orders as well as the value of the Dlog Pad\'{e} extrapolation when biasing with the known critical exponent of the 3d Ising universality class (horizontal dashed cyan line).}
  \label{fig:SM_pCUT_Gap}
\end{figure}

We performed a full graph decomposition on the triangular lattice and we focused on the 0qp and 1qp sector. The 0qp block contains solely the ground-state energy per site $e_0$, which we determined up to order 12 in $J$ and reads
\begin{eqnarray}
\label{Eq:e0_series}
 e_0 &=& -1 - \frac{3}{2}J^2- 3J^3 - \frac{309}{16} J^4 - \frac{405}{4} J^5  - \frac{83649}{128} J^6 - \frac{1128897}{256} J^7 - \frac{65234649}{2048}J^8 - \frac{490430787}{2048}J^9 \nonumber\\
 && - \frac{2103309861016488116883}{1125899906842624} J^{10}
 - \frac{16870782098514589885931}{1125899906842624}J^{11} -\frac{138603836752672268431157}{1125899906842624}J^{12}\quad .
\end{eqnarray}
The 1qp block represents a one-particle hopping Hamiltonian on the triangular lattice for a single excitation with $m=0$. It can be diagonalized by a Fourier transformation which yields the one-particle dispersion $\omega(\vec{k})$. The minimum of the dispersion corresponds to the one-particle gap $\Delta$, which is located at $\vec{k}=0$ for ferromagnetic interactions. We again reached order 12 in $J$ for this quantity. The series is given by 
 \begin{eqnarray}
\label{Eq:gap_series}
 \Delta &=& 1 - 6J - 15 J^2 - \frac{147}{2}J^3 - \frac{3705}{8}J^4 - \frac{103203}{32}J^5 - \frac{397611}{16} J^6 - \frac{14037072288726122493}{70368744177664} J^7\nonumber\\
        && - \frac{22635432824258351057353}{13510798882111488} J^8 - \frac{6232016455615499467244959}{432345564227567616} J^9 - \frac{9161896774402989476221315}{72057594037927936} J^{10}\nonumber\\
        && - \frac{5132573391520145132731581}{4503599627370496} J^{11}
           - \frac{373843963970302121493464787}{36028797018963968} J^{12}\quad .
\end{eqnarray}
We use Dlog Pad\'{e} extrapolation to locate the quantum critical point $(J/h)_{\rm c}^{\rm pCUT}$ between the polarized and the ordered phase of the spin-1 transverse-field Ising model on the triangular lattice. The corresponding results are displayed in Fig.~\ref{fig:SM_pCUT_Gap}. In addition, we biased the extrapolation with the known critical exponent $z\nu=0.6299$ of the 3d Ising universality class \cite{Kos2016} which yields $(J/h)_{\rm c}^{\rm pCUT,bias}=0.1899$. Overall, this yields the estimate $(J/h)_{\rm c}^{\rm pCUT}=0.1898(1)$ taking into account the biased value $(J/h)_{\rm c}^{\rm pCUT,bias}$ and the Dlog Pad\'{e} extrapolations of the highest order.

ical details, but directly give the physical results. 

\begin{figure}[ht]
  \centering \includegraphics[width=0.7\textwidth]{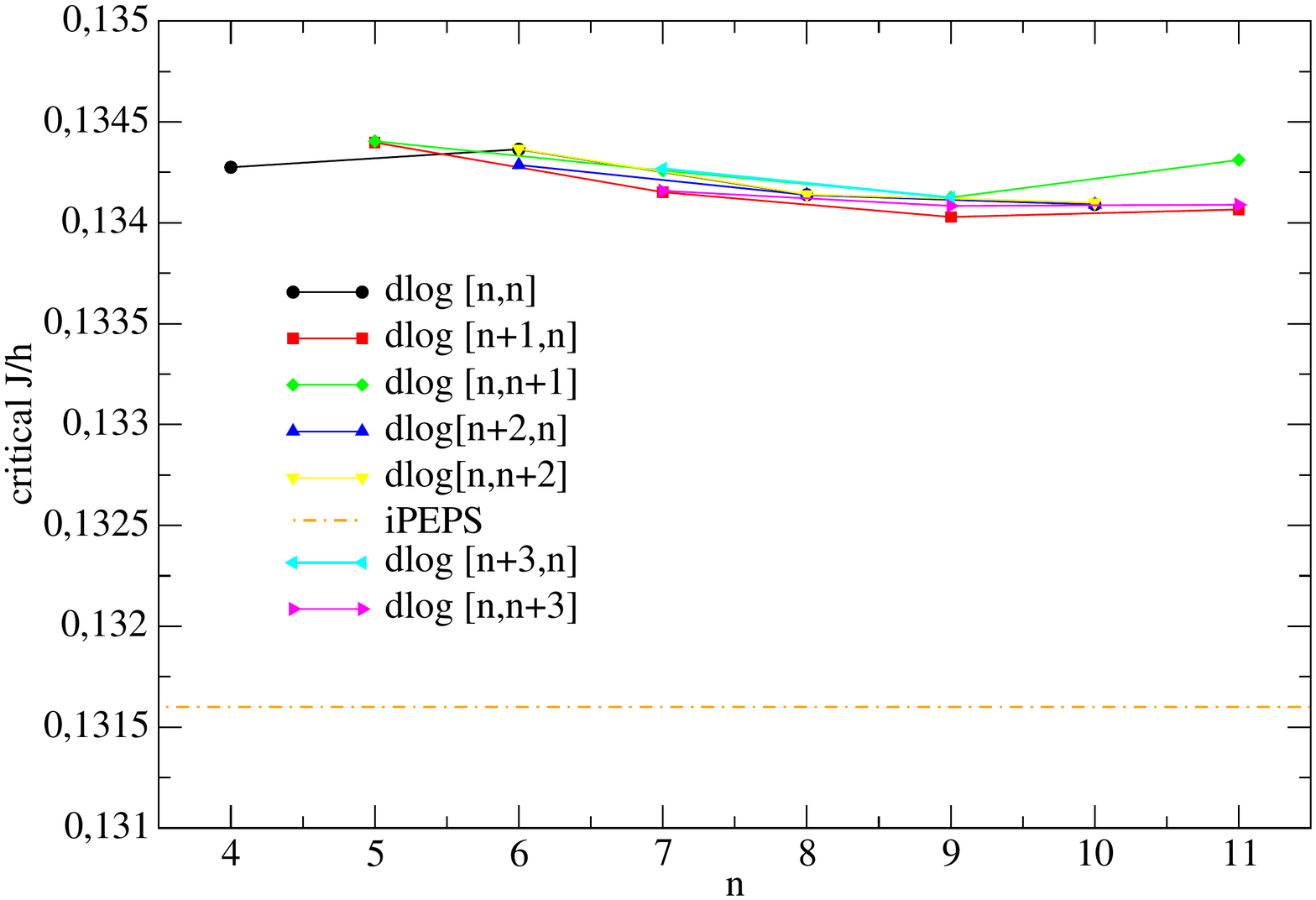}
  \caption{Critical point $(J/h)_{\rm c}^{\rm pCUT}$ of the spin-1 transverse-field Ising model on the stacked triangular lattice as a function of the perturbative order $n$ from Dlog Pad\'{e} extrapolation of the pCUT gap $\Delta$. Connected symbols represent a family of Dlog Pad\'{e} extrapolants $[L,M]$ with $L-M$ fixed. This results in the estimate $(J/h)_{\rm c}^{\rm pCUT}=0.13415(15)$, where the uncertainty reflects the difference of the families in the highest orders.}
  \label{fig:SM_pCUT_Gap2}
\end{figure}

The 0qp block contains solely the ground-state energy per site $e_0$, which we determined up to order 12 in $J$ and reads
\begin{eqnarray}
\label{Eq:e0_series2}
 e_0 &=& -1 -2J^2 -3J^3 -\frac{147}{4}J^4 -\frac{747}{4}J^5 -\frac{119763}{64}J^6 -\frac{3905067}{256}J^7 -\frac{20045217}{128}J^8 -\frac{3179261625}{2048}J^9\nonumber\\
 && -\frac{9473072468881078311871}{562949953421312}J^{10} -\frac{206653325264800591168511}{1125899906842624}J^{11}-\frac{1175460839841003872568547}{562949953421312}J^{12}\quad .
\end{eqnarray}
The 1qp block represents a one-particle hopping Hamiltonian on the 3d stacked triangular lattice for a single excitation with $m=0$. The minimum of the dispersion corresponds to the one-particle gap $\Delta$, which is again located at $\vec{k}=0$ for ferromagnetic interactions. We again reached order 12 in $J$ for this quantity. The series is given by 
 \begin{eqnarray}
\label{Eq:gap_series2}
 \Delta &=& 1-8J -28J^2 - 201J^3 - \frac{3669}{2}J^4 - \frac{300739}{16}J^5 -\frac{3343611}{16}J^6 - \frac{170834218382056751093}{70368744177664}J^7 \nonumber\\
&& - \frac{791727296122038217927439}{27021597764222976}J^8 - \frac{313304804862239506988778973}{864691128455135232}J^9- \frac{164976204678233268578164703}{36028797018963968}J^{10}\nonumber\\
&& - \frac{2117921992693265112120881413}{36028797018963968}J^{11}-\frac{6892798613801124481221437949}{9007199254740992}J^{12}\quad .
\end{eqnarray}
We use Dlog Pad\'{e} extrapolation to locate the quantum critical point $(J/h)_{\rm c}^{\rm pCUT}$ between the polarized and the ordered phase of the spin-1 transverse-field Ising model on the stacked triangular lattice. The corresponding results are displayed in Fig.~\ref{fig:SM_pCUT_Gap2}.  Overall, this yields the estimate $(J/h)_{\rm c}^{\rm pCUT}=0.13415(15)$ taking into account the Dlog Pad\'{e} extrapolations of the highest order.

\section{Exact calculations for the Mott phases of the Bose-Hubbard model on the Triangular lattice}
In the following we give some information on the exact analytical properties of the empty and completely filled Mott phases of the Bose-Hubbard model on the triangular lattice.

\subsection{Hard-Core Bose-Hubbard model on the Triangular lattice}
In the hardcore limit the local occupation numbers of the Bose-Hubbard model are restricted to zero and one. As a consequence, besides a superfluid phase, there are two Mott phases which are related exactly by particle-hole symmetry. For both Mott phases the ground state corresponds to exact product states which are given by $|0^{n=0}\rangle \equiv  |0\ldots 0 \rangle $ ($|0^{n=1}\rangle \equiv  |1\ldots 1 \rangle$) with ground-state energy $E_0^{n=0}=0$ ($E_0^{n=1}=-\mu N$) for density $n=0$ ($n=1$). As a consequence, the elementary gap of both Mott phases can be calculated exactly, since it reduces to a nearest-neighbor hopping problem of a single particle (single hole) on the triangular lattice which can be diagonalized by Fourier transformation. Specifically, we define a single-particle state  and a single-hole state on site $i$ by
\begin{eqnarray}
 |{\rm 1p}^{n=0},i\rangle &\equiv&  |0\ldots 0 1_i 0 \ldots 0 \rangle\\
 |{\rm 1h}^{n=1},i\rangle &\equiv&  |1\ldots 1 0_i 1 \ldots 1 \rangle\quad .
\end{eqnarray}
The kinetic term of the Bose-Hubbard model induces a nearest-neighbor hopping of the particle and the hole, respectively. The dispersion of both excitations is then given by 
\begin{eqnarray}
 \omega^{\rm 1p}(\vec{k}) &= &  -\mu -2t\left( cos(k_1)+cos(k_2)+cos(k_1-k_2)\right)\\
 \omega^{\rm 1h}(\vec{k}) &= &  +\mu -2t\left( cos(k_1)+cos(k_2)+cos(k_1-k_2)\right)\quad ,
\end{eqnarray}
where $\vec{k}=(k_1,k_2)$ denotes the two-dimensional wave-vector. The gaps are located at $\vec{k}=0$ so that
\begin{eqnarray}
 \Delta^{\rm 1p} &= &  -\mu -6t \\
 \Delta^{\rm 1h} &= &  +\mu -6t\quad .
\end{eqnarray}
These gaps close therefore exactly at $\mu/t=\pm 6$.\\

\subsection{Soft-Core Bose-Hubbard model on the Triangular lattice}

In the soft-core limit the local occupation numbers of the Bose-Hubbard model are restricted to zero, one, and two. As a consequence, besides a superfluid phase, there are three Mott phases. For the empty and completely filled Mott phase the ground state corresponds again to exact product states which are given by $|0^{n=0}\rangle \equiv  |0\ldots 0 \rangle $ ($|0^{n=2}\rangle \equiv  |2\ldots 2 \rangle$) with ground-state energy $E_0^{n=0}=0$ ($E_0^{n=2}=(U-2\mu) N$) for density $n=0$ ($n=2$). As a consequence, the elementary gap of these two Mott phases can again be calculated exactly. Specifically, we define a single-particle state and a single-hole state on site $i$ by
\begin{eqnarray}
 |{\rm 1p}^{n=0},i\rangle &\equiv&  |0\ldots 0 1_i 0 \ldots 0 \rangle\\
 |{\rm 1h}^{n=2},i\rangle &\equiv&  |2\ldots 2 1_i 2 \ldots 2 \rangle\quad .
\end{eqnarray}
The dispersion of both excitations is then given by 
\begin{eqnarray}
 \omega^{\rm 1p}(\vec{k}) &= &  -\mu -2t\left( cos(k_1)+cos(k_2)+cos(k_1-k_2)\right)\\
 \omega^{\rm 1h}(\vec{k}) &= &  +\mu -4t\left( cos(k_1)+cos(k_2)+cos(k_1-k_2)\right) -U\quad ,
\end{eqnarray}
where $\vec{k}=(k_1,k_2)$ denotes the two-dimensional wave-vector. The gaps are located at $\vec{k}=0$ so that
\begin{eqnarray}
 \Delta^{\rm 1p} &= &  -\mu -6t \\
 \Delta^{\rm 1h} &= &  +\mu -12t-U\quad .
\end{eqnarray}

\bibliography{references}{}
\bibliographystyle{apsrev4-1}